\documentclass[twocolumn,showpacs,showkeywords,prb]{revtex4}
\usepackage{graphicx}
\usepackage{dcolumn}
\usepackage{amsmath}
\usepackage{latexsym}
\begin{document}
\title
{COW test of the weak equivalence principle: A low-energy window to look into the noncommutative structure of space-time?}
\author{Anirban Saha}
\email{anirban@iucaa.ernet.in }
\altaffiliation{Visiting Associate in Inter University Centre for Astronomy
and Astrophysics, Pune, India}
\affiliation{Department of Physics, West Bengal State University,Barasat,North 24 Paraganas, West Bengal, India\\ }

\begin{abstract}
{\noindent } 
We construct the quantum mechanical model of the COW experiment assuming that the underlying space time has a granular structure, described by a canonical noncommutative algebra of coordinates $x^{\mu}$. The time-space sector of the algebra is shown to add a mass-dependent contribution to the gravitational acceleration felt by neutron deBrogli waves measured in a COW experiment. This makes time-space noncommutativity a potential candidate for an apparent violation of WEP even if the ratio of the inertial mass $m_{i}$ and gravitational mass $m_{g}$ is a universal constant. The latest experimental result based on COW principle is shown to place an upper-bound several orders of magnitude stronger than the existing one on the time-space noncommutative parameter. We argue that the evidence of NC structure of space-time may be found if the COW-type experiment can be repeated with several particle species.

\end{abstract}
%Starting form a field theoretic frame work we arive at a commutative equivalent description of the basic COW experimental setup where the underlying space time is assumed to follow a noncommutative geometry. It is shown that the noncommutativity in the time-space sector will lead to a mass-dependent gravitational acceleration and therefore an aperant violation of the equivalence principle even if the ratio between the gravitational and inartial mass is an universal constant.
\maketitle

	At the Planck scale the space-time is thought to have a granular structure that can be described by a noncommutative (NC) geometry with space-time coordinates $x^{\mu}$ satisfying the algebra
\begin{equation}
\left[x^{\mu}, x^{\nu}\right] = i \Theta^{\mu \nu}
\label{ncgometry}
\end{equation}
where $\Theta^{\mu \nu}$ is a constant anti-symmetric tensor. This idea of a NC space-time has gained interest in the recent past when it was commonly realized that the low energy effective theory of D-brane in the background of NS-NS B field lives on noncommutative space \cite{sw, sz}. In the brane world scenerio \cite{Antoni}, our spacetime may be the world volume of a D-brane, and thus can be described by noncommutative geometry (\ref{ncgometry}). Also, from the physical perspective it has long been suggested that in the Gedanken experiment of localizing events in a space-time with  Planck scale resolution, a sharp localization induces an uncertainty in the space-time coordinates which can be naturally described by the noncommutative geometry (\ref{ncgometry}) \cite{Dop, Alu}. Although effects of such a NC structure of space-time may appear near the string/Planckian scale, it is hoped that some low energy relics of such effects may exist and their phenomenology can be explored at the level of quantum mechanics (QM) \cite{pmh, cst, bcsgas, Nair, sahasg, sahagan, bert0}. 

	The structure of space-time is best revealed through gravitational interaction. In fact, the central idea of Einstein's general theory of relativity (GTR) is founded on the interpretation of gravity as a property of space-time, namely its curvature. This interpretation is largely based upon the Weak Equivalence Principle (WEP)which has its experimental foundation in the universality of free fall (UFF) that demands a universally constant ratio $\frac{m_{i}}{m_{g}} = \alpha$ between the inertial mass $m_{i}$ and gravitational mass $m_{g}$, both appearing in the classical equation of motion  
\begin{eqnarray}
%\frac{d^{2} x}{dt^{2}}
\ddot{x} =  \frac{m_{g}}{m_{i}}  g = \frac{g}{\alpha}  = g^{\prime}
\label{uff}
\end{eqnarray}
of a freely falling ``point like" particle immersed in the nearly homogeneous local gravitational field $g = \frac{G M_{E}}{R_{E}{}^{2}}$ caused by Earth's mass $M_{E}$. Here we have ignored the nominal height from ground level $h$ with respect to the Earth's radius $R_{E}$. The effect of this gravitational field $g$ is the gravitational acceleration of the particle $\ddot{x} = g^{\prime} = \frac{g}{\alpha}$ which, if $\alpha$ indeed is a universal constant and does not vary from one particle species to another, is same for all kind of material particles, as the UFF demands. 

	Curiously, most theoretical attempts to connect GTR to standard model allows for violation of WEP \cite{vio1, vio2, vio3, vio4, vio5, vio6, vio7} and therefore it has a long and persistent history of experimental tests of various kind so that insight into some alternative/modified version of GTR may be obtained. 

	In experimental tests of WEP with macroscopic objects we look for species-dependent value of the gravitational acceleration $g^{\prime}$ caused by change in the value of $\alpha$ for different particle species. 
%which measures the difference in acceleration of 
In the E\"{o}tv\"{o}s-type experiments possible violations are parametrised by the E\"{o}tv\"{o}s factor, $\eta \left(A,B\right) = \frac{\delta g^{\prime}\left(A,B\right)}{g^{\prime}_{average}\left(A,B\right)} = 2\frac{\ddot{x}\left(A\right) - \ddot{x}\left(B\right)} {\ddot{x}\left(A\right) + \ddot{x}\left(B\right)}$ for two macroscopic test masses made of materials A and B. Currently the lowest bound is reached for the elements Beryllium and Titanium, using rotating torsion balances \cite{torsion}, $\eta\left(\rm{Be, Ti}\right) < 2.1 \times 10^{-13}$. Future tests like MICROSCOPE \cite{MICROSCOPE} %\footnote{``MICRO-Satellite \`{a} tra\^{i}n\'{e}e Compens\'{e}e pour l'Observation du Principe d'Equivalence",}
to be launched in 2014 aim at a lower bound of $10^{-15}$. In the atomic/subatomic regime improvements of earlier E\"{o}tv\"{o}s-type experiments, by Dicke {\it et al} \footnote{They compared acceleration of PbC1 (neutron-proton ratio R=1.45) and of Cu (R= 1.19).}
in 1961 concluded that neutrons and protons in nuclei experience the same gravitational acceleration $g^{\prime}$ within about $2\times 10^{-9} g^{\prime}$ \cite{Dicke}. That a free neutron experiences the same $g^{\prime}$ it experiences within a nucleus was experimentally confirmed \cite{freeneutron} in 1965 by measuring $g^{\prime}$ from the difference of fall of two well-collimated beams of high and low velocity neutrons while traversing a long evacuated horizontal flight path.
A comparison of neutron scattering lengths, with measurement techniques both dependent \footnote{Slow neutrons reflected from liquid mirrors after having fallen a hight.}
and independent \footnote{By transmission measurements on liquid prob.}
of gravity, also leads to a verification of the WEP \cite{scattering} in 1976. These results, though obtained for free neutrons behaving as matter waves, are still a consequence of their classical parabolic path under gravity as required by the correspondence principle and hence no quantum features are involved. 

	During 1974 to 1979, when Colella, Overhauser and Werner (COW), in a series of experiments \cite{cow} demonstrated the validity of WEP using gravitationally induced quantum-mechanical phase shift in the interference between coherently split and separated neutron de Broglie waves at the 2MW University of Michigan Reactor, the validity of equivalence principle in the so called ``quantum limit" was claimed to have been examined. The verification was complimented in 1983 by repeating the experiment in an accelerated interferometer where gravitational effects are compensated \cite{noninertial}. This established that the Schr\"{o}dinger equation in an accelerated frame predicts a phase shift which agrees with observation as assumed earlier by COW \cite{COW2} for the validity of strong equivalence principle in the quantum limit. Since then, the equivalence principle in the quantum limit has been verified, time and again, with ever increasing accuracy.

 Given the roll of WEP in attributing gravity as a property of the space-time, a COW test can be regarded as a test of the space-time property at the quantum level. Therefore, it will not be surprising if some trace of the space-time structure at the Plank scale resolution manifest itself, even in the low energy regime where quantum mechanical tests of WEP are currently being performed.

	In this letter we therefore construct the quantum mechanical theory describing the basic COW experiment with the assumption that the underlying space-time we live in follows a NC geometry described by (\ref{ncgometry}). Our motivation is to investigate if some manifestation of this NC structure shows up in the observable results. Specifically, we work out the gravity-induced phase-shift which shows a leading order NC contribution. It is argued that this NC term will lead to an apparent violation of WEP in COW-type test data. In the latest experiments based on COW principle \cite{lit} the WEP is verified to $1\%$ precision level. This result is employed to put an upper-bound on the NC parameter which turns out to be stronger than the existing bound \cite{ani}.% estimated using the GRANIT experimental results \cite{nes1, nes2, nes3}. 
We also put forward a suggestion to trace this apparent violation of the WEP to its' NC origin if such COW-type experiments can be performed with different atomic/subatomic particles. This can serve as an evidence of the NC structure of space-time. 

We start by discussing how to introduce the NC space-time structure in the system. Since in QM space and time could not be treated on an equal footing, we impose the geometry (\ref{ncgometry}) at a field theoretic level and eventually reduce the theory to quantum mechanics \footnote{This is a reasonable starting point since single particle quantum mechanics can be viewed as the one-particle sector of quantum field theory in the very weakly coupled limit where the field equations are essentially obeyed by the Schr\"{o}dinger wave function \cite{Nair, pmh, bcsgas}.}. This allows us to examine the effect of the whole sector of space-time noncommutativity in an effective noncommutative quantum mechanical (NCQM) theory. Owing to the extreme smallness of the NC parameters the current/near future experiments can only hope to detect the first order NC effects. Since it has been demonstrated in various formulations of NC gravity \cite{grav1} that the leading NC correction in the gravity sector is second order we can safely assume the Newtonian gravitational field $g$ remains unaltered for all practical purpose.

	The NC Schr\"{o}dinger field theory describing cold neutron beams in Earth's gravitational field (along the $x$-axis) in a vertical $xy$ ($i = 1, 2$) plane is 
\begin{eqnarray} 
\hat S = \int d^{2}x dt  \hspace{1.0mm} \hat \psi^{\dag}\star \left[i \hbar \partial_{0} + \frac{{\hbar}^{2}}{2m_{i}} \partial_{i}\partial_{i} - m_{g} g\hat{x} \right] \star \hat \psi
\label{NCaction} 
\end{eqnarray}
Since there is no direct way to relate the physical observables to the NC operators in (\ref{NCaction}) we consider the NC fields $\hat \psi$ as functions in the deformed phase space where ordinary product is replaced by the star product \cite{sw, bcsgas} which, for two fields $\hat \phi(x)$ and $\hat \psi(x)$, is given by 
\begin{equation}
\hat \phi(x) \star \hat \psi(x) = \left(\hat \phi \star \hat \psi \right)(x) = e^{\frac{i}{2}
\theta^{\alpha\beta}\partial_{\alpha}\partial^{'}_{\beta}}
  \hat \phi (x) \hat \psi(x^{'})\big{|}_{x^{'}=x.}
\label{star}
\end{equation}
Due to the linear form of the gravitational potential in action (\ref{NCaction}), expanding the star product and expressing  everything in terms of commutative variables only gives corrections to first order in the NC parameters and all the higher order terms vanish. 
This leads to an equivalent commutative description of the original NC model in terms of the non-canonical action
\begin{eqnarray} 
\hat S = \int  d^{2}x dt  \hspace{1.0mm} \psi^{\dag}\hspace{-4.0mm}  && \left[i \hbar \left( 1  - \frac{\eta}{2\hbar} m_{g} g  \right) \partial_{t}  +  \frac{{\hbar}^{2}}{2m_{i}} \partial_{i}{}^{2}\right. \nonumber\\
&&\left. - m_{g} g x  - \frac{i}{2} m_{g} g \theta \partial_{y} \right] \psi
\label{Caction} 
\end{eqnarray} 
where NC effect is manifest by the presence of NC parameter $\Theta^{10} = \eta$ among time and spatial directions . The term  with spatial NC parameter $\Theta^{12} =\theta$ and first derivative $\partial_{y}$ can be absorbed in the $\partial_{y}{}^{2}$ and is therefore inconsequential.

 We use a physically irrelevant rescaling\footnote{Since the experimental setup is confined to a small region of space-time where the local gravitational field $g$ is essentially constant, this rescaling amounts to multiplying the field variable by a constant.} of the fields 
%\begin{eqnarray}
$\psi \mapsto \tilde{\psi} = \sqrt{\left(1 - \frac{\eta}{2 \hbar} m_{g} g\right)} \psi$
%\label{scal1}
%\end{eqnarray}
to recast this non-canonical form of action with a conventionally normalized kinetic term such that the fields evolves in a  canonical manner. This leads to 
\begin{eqnarray} 
\hat S = \int  d^{2}x dt \hspace{1.0mm} \tilde{\psi}^{\dag}\hspace{-4.0mm}  && \left[i \hbar \partial_{t}  +  \frac{{\hbar}^{2}}{2m_{i}\left(1 - \frac{\eta}{2 \hbar} m_{g} g\right)} \partial_{i}{}^{2} \right.\nonumber  \\ 
&& \left.- \frac{m_{g} g x}{\left( 1  -  \frac{ \eta m_{g} g }{2\hbar} \right) } \right] \tilde{\psi}
\label{Caction1} 
\end{eqnarray} 
Comparing with the standard Schr\"{o}dinger action we can immidiately read off the observed inertial mass as $ \tilde {m}_{i} = 2m_{i}\left(1 - \frac{\eta}{2 \hbar} m_{g} g\right)$. Assuming the NC effect to be very small the interaction can be written in terms of this observed inartial mass $\tilde{m}_{i}$ as
\begin{eqnarray} 
\frac{m_{g} g x}{\left( 1  -   \frac{\eta m_{g} g }{2\hbar} \right) }  
%=  \frac{m_{i} g^{\prime} x}{ \left( 1 - \frac{\eta m_{i} g^{\prime} }{2\hbar} \right)} 
= \tilde{m}_{i} g^{\prime} x \left( 1  +  \frac{\eta \tilde{m}_{i} g^{\prime} }{\hbar} \right)
\label{mtilde} 
\end{eqnarray} 
where we have used equation(\ref{uff}) to replace $m_{g} g $ with $m_{i} g^{\prime}$. 

Note that replacing $m_{g} g $ with $m_{i} g^{\prime}$ follows from the definition of gravitational acceleration $g^{\prime}$ for an individual particle, as in (\ref{uff}), and not from the assumption of WEP. WEP is required when we assume that such accelerations for two separate particle species are identical for same gravitational field $g$. 

The final form of the canonical action reads 
\begin{eqnarray} 
\hat S = \int  d^{2}x dt \hspace{1.0mm} \tilde{\psi}^{\dag}\hspace{-4.0mm}  && \left[i \hbar \partial_{t}  +  \frac{{\hbar}^{2}}{2\tilde{m}_{i}} \partial_{i}{}^{2} -  \tilde{m}_{i} g^{\prime} x \left( 1  +  \frac{\eta \tilde{m}_{i} g^{\prime} }{\hbar} \right)\right] \tilde{\psi} \nonumber\\
\label{Caction2} 
\end{eqnarray} 
leading to the equation of motion 
\begin{eqnarray} 
i \hbar \partial_{t}  \tilde{\psi}  = -  \left[\frac{{\hbar}^{2}}{2\tilde{m}_{i}} \partial_{i}{}^{2} + \tilde{m}_{i} g^{\prime} x \left( 1  +  \frac{\eta \tilde{m}_{i} g^{\prime} }{\hbar} \right) \right] \tilde{\psi} 
\label{eqm} 
\end{eqnarray} 
that can be considered at the level of quantum mechanics with $ \tilde{\psi}$ interpreted as the Schr\"{o}dinger wave function. Equation (\ref{eqm}) describes the NCQM of a freely falling neutron in earth's gravity in terms of commutative variables. We can readily derive the Ehrenfest relations 
\begin{eqnarray}
\frac{d}{dt}<x> &=& \frac{<p>}{\tilde{m}_{i}}  \label{velocity}\\ 
\frac{d^{2}}{dt^{2}}<x> &=& g^{\prime}\left( 1 + \frac{\eta \tilde{m}_{i} g^{\prime} }{\hbar} \right) = \tilde{g}^{\prime} \label{acceleration}
\end{eqnarray}
for the average velocity and acceleration of the neutrons. Thus, though representing an NC system, this Schr\"{o}dinger equation (\ref{eqm}) behaves similar to that in ordinary/commutative space. However, the two crucial differences with the commutative result are 
\begin{enumerate}
\item the appearance of observed inertial mass of the neutron $\tilde{m}_{i}$ in the average momentum (\ref{velocity})  \\
and 
\item the observed gravitational acceleration $\tilde{g}^{\prime} $ in (\ref{acceleration}) experienced by a quantum mechanically behaving system, namely the neutron, is now mass-dependent due to the NC structure of space-time. 
\end{enumerate}
Note that contrary to the common expectation that Ehrenfest theorem will lead to results mimicking classical behaviour i.e. a quantum mechanical wave packet will move, on an average, along a classical particle trajectory subject to the applied potential \cite{saku}, here we have a observable quantum mechanical effect that is not washed out by the averaging process and shows up as a deviation from the classical trajectory. That this effect is of NCQM origin is established by the explicit appearance of the ratio $\frac{\eta}{\hbar}$. 

In a COW-type experimental setting the gravitational potential is much smaller than the total energy of the neutrons and we can calculate the gravity induced phase-shift from (\ref{eqm}) by the semi-classical prescription of matter-wave interferometry \cite{berman, rmp} 
\begin{eqnarray}
\Delta \varphi_{grav} = - \frac{1}{\hbar} \tilde{m}_{i} \tilde{g}^{\prime} \left( l_{1} \sin \phi \right) \Delta t 
\label{phase_shift}
\end{eqnarray}
where $\phi$ is the tilt angle between the plane containing the coherently splitted neutron beams and the horizontal plane, giving rise to an effective height $l_{1}\sin \phi $ of one of the neutron beam paths with respect to the other.
Since the effective potential is time-independent here we can use the paraxial approximation to compute 
\begin{eqnarray}
\Delta t = {l_{2}}/{\frac{d}{dt}<x>} = \frac{l_{2} \tilde{m}_{i}\lambda_{0}}{h} 
\label{deltat}
\end{eqnarray}
where $\lambda_{0} = h/<p>$ is the laboratory neutron deBrogli wavelength corresponding to the average neutron momentum $<p>$ in (\ref{velocity}). Combining (\ref{phase_shift}) and (\ref{deltat}) we find 
\begin{eqnarray}
\Delta \varphi_{grav} = - \frac{A \sin \phi}{ 2 \pi \hbar^{2}} \lambda_{0} \tilde{m}_{i}{}^{2} \tilde{g}^{\prime}
\label{phase}
\end{eqnarray}
where $A = l_{1} l_{2}$ is the area enclosed by the interfering beams. This phase difference depends on the mass-dependent $\tilde{g}^{\prime}$. 

Comparing this theoretical prediction (\ref{phase}) with the experimentally measured gravity induced phase-shift one can obtain the {\it quantum mechanically observed} gravitational acceleration $\tilde{g}^{\prime}\left({\rm n}\right)$ felt by a neutron. We intend to stress the quantum mechanical nature of the observation because phase-shift is a quantum phenomena and it is only in the quantum regime that any NC effect will be picked up. This data, when confronted with local classical gravitational acceleration $g^{\prime}$ measured with macroscopic bodies where no NC effect is possible, will exhibit a discrepancy given by 
\begin{eqnarray}
\frac{\delta g}{g_{{\rm av}}} =   \frac{\tilde{g}^{\prime}\left({\rm n}\right) - g^{\prime}}{g_{{\rm av}}} =  \frac{{g}^{\prime}\left({\rm n}\right) - g^{\prime}}{g_{{\rm av}}} + \frac{\eta \tilde{m}_{i} \left(g^{\prime}\left({\rm n}\right)\right)^{2} }{\hbar g_{{\rm av}}}
\label{discrepancy}
\end{eqnarray}
Here ${g}^{\prime}\left({\rm n}\right) = \frac{g}{\alpha\left({\rm n}\right)}$ is the acceleration the neutron would feel due to Earth's gravitational field $g$ if our space-time followed the ordinary Hisenberg algebra
%$\left[x^{\mu}, x^{\nu}\right] = 0; \left[p^{\mu}, p^{\nu}\right] = 0; \left[x^{\mu}, p^{\nu}\right] = i\hbar \delta^{\mu \nu}$ 
instead of the NC algebra (\ref{ncgometry}). The first term signifies the violation of the WEP, if any, caused by the non-universality of $\alpha$, i.e. $\alpha\left({\rm n}\right) \ne \alpha\left({\rm macroscopic}\right)$ and the second term arise as an effect of the NC structure of space-time showing an apparent violation even if $\alpha$ in (\ref{uff}) is a universal constant. This sets a limitation on the accuracy to which WEP can be verified at the quantum limit by COW  experiments on ultra-cold neutrons. 

Assuming that WEP holds up to a higher accuracy level than where the NC effect makes its presence felt, an upper-bound can be set on the NC parameter $\eta$ using available experimental data. The first contribution then vanishes and any discrepancy is only due to the second term. In recent years, Littrell et al \cite{lit} used nearly harmonic pairs of neutron wavelengths\footnote{The harmonic pairs of neutron wavelengths were used to compensate for effects due to the distortion of the interferometer as it was tilted about the incident beam direction.} with perfect silicon crystal interferometers and showed a discrepancy of 1\% in the observed gravity-induced phase-shift with the theoretical value. Using this result the upper bound on $\eta$ is found to be
\begin{eqnarray}
|\eta| \lesssim 6.4248 \times 10^{-13} {\rm m}{}^{2}
\label{ub}
\end{eqnarray}
This bound is stronger than the earlier bound on the time-space NC parameter \cite{ani} given by 
\begin{eqnarray} 
|\eta| & \lesssim & 2.83\times 10^{-9}\ \mathrm{m^{2}}
\label{eta_bounds_1}
\end{eqnarray}
estimated using the GRANIT experimental results \cite{nes1, nes2, nes3}. 

In principle the apparent violation due to NC effect should be identifiable if the COW-type experiments can be performed with different atomic/subatomic particle species. With the first term vanishing/negligible in (\ref{discrepancy}), the discrepancy for different species will vary linearly with their masses and the slope $\frac{\eta g^{\prime}}{\hbar}$ will give the absolute value of the NC parameter. Such a linear variation of discrepancy with particle mass, if indeed observed, will serve  to establish the granular structure of the space-time we live in. Of course this holds only if any true violation due to non-universality of $\alpha$ occurs beyond the accuracy level where the NC effect starts affecting the data. In the best case scenario the COW-type experiments and its other variants such as atom-interferometer based on fountain of laser-cooled atoms \cite{nat}, may open a low-energy ``window" to reveal the noncommutative structure of space-time.

The author would like to thank Patrick DasGupta, Sunandan Gangopadhyay, Bibhash Bhattacharya and Pradip Mukherjee for useful discussion and enlightening comments.

\end{document}